\documentclass[aps,prl,amsmath,amssymb,twocolumn,superscriptaddress]{revtex4-2}
\pdfoutput=1 
\usepackage[usenames]{xcolor}
\usepackage[pdftex]{graphicx}
\usepackage{amsmath, amstext, amssymb, amsfonts, amsxtra}
\usepackage{xspace}
\usepackage{blindtext}
\usepackage{braket}
\usepackage{dsfont}
\usepackage{placeins}
\usepackage{nicefrac}
\usepackage{here}
\usepackage[colorlinks=true,linkcolor=black,citecolor=blue,urlcolor=black]{hyperref}
\usepackage{siunitx}

\begin{document}
\title{Non-Abelian vibron dynamics in trapped-ion arrays}
\author{L. Timm}
\email[]{lars.timm@itp.uni-hannover.de}
\affiliation{Institut f\"ur Theoretische Physik, Leibniz Universit\"at Hannover, Appelstr. 2, 30167 Hannover, Germany}
\author{H. Weimer}
\affiliation{Institut f\"ur Theoretische Physik, Leibniz Universit\"at Hannover, Appelstr. 2, 30167 Hannover, Germany}
\affiliation{Institut f\"ur Theoretische Physik, Technische Universit\"at Berlin, Hardenbergstr. 36,
10623 Berlin, Germany}
\author{L. Santos}
\affiliation{Institut f\"ur Theoretische Physik, Leibniz Universit\"at Hannover, Appelstr. 2, 30167 Hannover, Germany}
%\author{T. E. Mehlst\"aubler}
%\affiliation{Physikalisch-Technische Bundesanstalt, Bundesallee 100, 38116 Braunschweig, Germany}

\date{\today}

\begin{abstract}
Trapped-ion arrays offer interesting possibilities for quantum simulation. We show that a proper arrangement of elliptical micro-traps combined with the external driving of the micro-trap frequencies allows, without the need of any precise fine-tuning, for the robust realization of non-Abelian vibron dynamics. 
We show that this non-Abelian nature may be readily probed experimentally in a simple plaquette arrangement. This study opens interesting perspectives for the study of non-Abelian spin-orbit coupling with motional excitations in two- and three-dimensional ion arrays.
\end{abstract}

\maketitle

%%%%%%%%%%%%%%%%%%%%%%%%%%%%%%%%%%%%%%%%%%%%%%%%%%%%%%%%%%%%%%%%%%%%%%%%%%%%%%%%%%%%

% INTRODUCTION

%\section{Introduction}

Systems composed of cold atoms and ions offer a platform for a variety of applications in quantum simulation \cite{lewenstein2007,haffner2008,bruzewicz2019}.
Today there exists a plethora of different approaches.
They range from neutral atoms trapped in optical lattices or magneto optical traps over ions in radiofrequency or optical traps to highly excited Rydberg atoms \cite{bloch2005,browaeys2020,drewsen2015,schaetz2017}.
With these different systems come different types of interaction that can be suitable for the simulation of the quantum many-body physics of a desired model, e.g. Hubbard type models with contact or long-range interactions \cite{hubbard1963,jaksch2005,chomaz2023}.

Special attention has attracted the realization of cold gases in the presence of synthetic gauge fields~\cite{jaksch2003,dalibard2011,goldman2014a,goldman2014,celi2014,dalibard2015,aidelsburger2016}, which present interesting properties such as fractal band structures and topological features, and offer a larger flexibility than other condensed-matter systems aiming at similar phenomena \cite{hofstadter1976,aidelsburger2011,cooper2019,braun2023,wintersperger2020}.
All reported schemes for the generation of synthetic fields make use of some form of external drive, either addressing one or more electronic transitions or employing Floquet techniques modulating the trapping of the particles \cite{shirley1965,haldane1988,weitenberg2021,eckardt2017}. Although initially these artificial fields were externally imposed, and hence static, recent years have witnessed a growing attention on density-dependent and dynamical gauge fields \cite{Goerg2019,lienhard2020,aidelsburger2022}. 

Coupling the motion of the particles in an optical lattice to a pseudospin degree of freedom may result in the  realization of non-Abelian gauge fields \cite{osterloh2005,ruseckas2005,hauke2012}.
Spin-orbit coupling has been successfully demonstrated for ultracold atoms in the continuum with the help of Raman laser beams \cite{lin2011a,Wang2012,Cheuk2012,Huang2016}. 
%galitski2013,zhai2015}.
While these realizations require the control of the electronic states of the neutral atoms, we encode in this paper the pseudo-spin in the motional excitations of ions in a trap array.
These excitations, denoted vibrons in the following, move via Coulomb interaction through an array of ions confined by micro-traps \cite{ramm2014,tamura2020}.
State-of-the-art experiments may controllably 
drive the local trap frequencies, and allow for a precise  engineering of the hopping of vibrons amongst the different ions of the array \cite{schmied2009,mielenz2016,gorman2018,hakelberg2019,kiefer2019}. It has been proposed that these capabilities could allow for the creation of an artificial Abelian gauge field that couples to the vibrons \cite{bermudez2011,bermudez2012}. In this paper, we show that a proper choice of the arrangement of the micro-traps combined with an external driving of the micro-trap frequencies allows for the robust realization of non-Abelian vibron dynamics, which may be readily probed in experiments using a simple-plaquette arrangement.

%%%%%%%%%%%%%%%%%%%%%%%%%%%%%%%%%%%%%%%%%%%%%%%%%%%%%%%

% FIGURE 1

\begin{figure}
\includegraphics[width = \columnwidth]{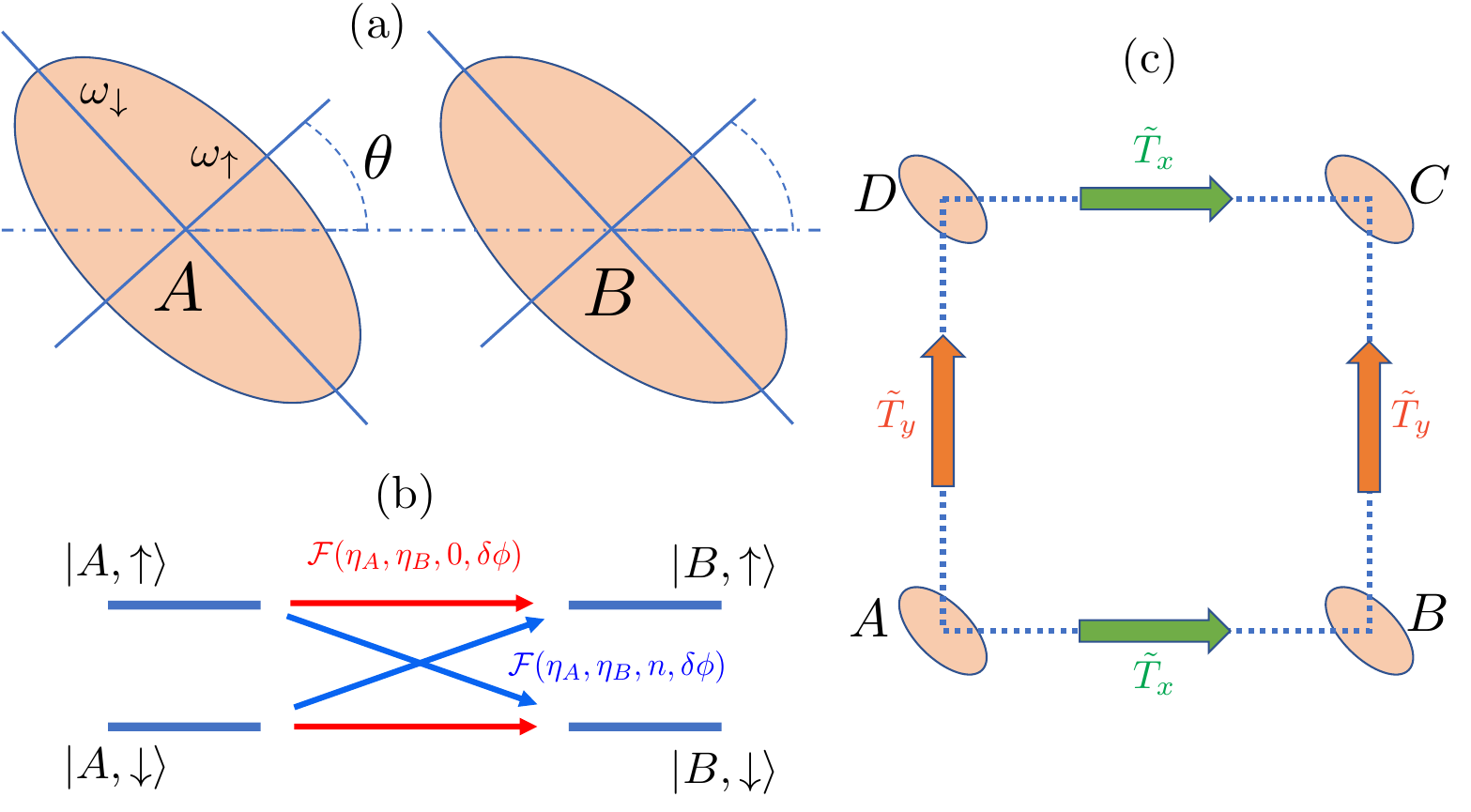}
\caption{(a) Scheme of two neighboring elliptical microtraps whose axes sustain an angle $\theta$ with the line joining their centers; 
(b) vibron states of neighboring ions coupled by diagonal and off-diagonal terms; (c) square plaquette of tilted microtraps discussed in the text.}
\label{fig:schemes}
\end{figure}

%%%%%%%%%%%%%%%%%%%%%%%%%%%%%%%%%%%%%%%%%%%%%%%%%%%%%%%

%%%%%%%%%%%%%%%%%%%%%%%%%%%%%%%%%%%%%%%%%%%%%%%%%%%%%%%%%%%%%%
%%% FORMALISM %%%

\paragraph{Floquet-engineered vibron dynamics.--} 
In order to introduce the main requirements for the engineering of  non-Abelian dynamics,  we first discuss the simple case of two ions, denoted $A$ and $B$, of mass $m$ and charge $e$, see Fig.~\ref{fig:schemes}(a). 
The ions are confined in separated harmonic micro-traps of frequencies  $\omega_{\uparrow}>\omega_\downarrow$ and trap axes $\vec e_{\uparrow}=\cos\theta\vec e_x+\sin\theta\vec e_y$, $\vec e_{\downarrow}=-\sin\theta\vec e_x+\cos\theta\vec e_y$. 
The trap centers $\vec R_{j=A,B}$ are aligned along the $x$ axis at a distance $d$. 
The two-ion system is well described the Hamiltonian:
\begin{align}
\!\hat H \!&=\! \sum_{j}\! \left [\frac{\hat p_j^2}{2m}\!+\!\frac{m}{2}\!\!\sum_{s=\uparrow,\downarrow}\!
\omega_s^2 \left ( \vec r_j \cdot \vec e_s \right )^2 \!\right ]\! 
\!+\! \frac{C_0}{ |d\vec e_x\!+\!\vec r_A\!-\!\vec r_B|}, \\
\end{align}
where $\vec r_j$ denotes the displacement of the $j$-th ion off the trap center $\vec R_j$, $\hat p_j$ is the corresponding momentum operator, and $C_0=e^2/4\pi\epsilon_0$, with $\epsilon_0$ the vacuum permittivity. 

Quantized ion oscillations around the trap centers~(called vibrons in the following) are created~(annihilated) by the operators $a_{js}^\dagger$~($a_{js}$). 
Expanding the Coulomb interaction for $|\vec r_A-\vec r_B|\ll d$, and assuming that $\hbar\omega_s \gg \frac{C_0l_s^2}{d_x^3}$, with $l_s$ the harmonic oscillator lengths, we obtain the Hamiltonian:
\begin{align}
H = \hbar \sum_{j,s}\omega_s a_{js}^\dag a_{js} + \hbar \sum_{s,s'} 
\left ( a_{Bs}^\dag T_{ss'} a_{As'} + \mathrm{H. c.} \right ),
\label{eq:H}
\end{align}
where 
\begin{align}
T(\theta)
&=\frac{C_0}{md^3}
\begin{pmatrix}
\frac{1-3\cos^2\theta}{\omega_\uparrow} & \frac{3\sin\theta\cos\theta}{\sqrt{\omega_\uparrow\omega_\downarrow}} \\ \frac{3\sin\theta\cos\theta}{\sqrt{\omega_\uparrow\omega_\downarrow}}  & \frac{1-3\sin^2\theta}{\omega_\downarrow}
\end{pmatrix}
\label{eq:T}
\end{align}
is the vibron-hopping matrix \cite{[{For a derivation of the vibron Hamiltonian see the Supplementary Material}]SMvibron}. 
The system hence emulates the Hubbard-like dynamics of pseudo-spin-$1/2$ bosons hopping between the two micro-traps, with the spin encoded in the two oscillation directions of the micro-traps. 
Note that 
\begin{align}
    \omega_\uparrow-\omega_\downarrow &\gg \frac{3C_0l_\uparrow l_\downarrow}{4d^3}\sin2\theta
    \label{eq:requirement}
\end{align} 
is necessary in order to guarantee that the local oscillator axes are only given by the microtrap potentials, independently of the Coulomb interaction with neighboring ions. 

The hopping matrix is diagonal if one of the micro-trap axes aligns with the line joining the micro-trap centers~($\theta=0$ or $\pi/2$). 
In that case, the pseudo-spin vibron components decouple, a situation encountered in trapped-ion arrays employed for quantum computation~\cite{cirac1995}. 
An angle $\theta\neq 0,\pi/2$ is hence required to get off-diagonal hopping terms, a necessary requirement to establish non-Abelian dynamics. 
However, off-diagonal coupling is off-resonant, due to condition~\eqref{eq:requirement}, and hence negligible. 
As a result, hopping would be effectively diagonal, precluding the study of non-Abelian dynamics.

Off-diagonal hopping may be however activated by means of Floquet techniques, assuming that the trap frequencies are periodically modulated with a driving frequency $\omega_d$:
\begin{align}
\omega_{i,s=\uparrow,\downarrow}(t)=\omega_s + \eta_{i}\omega_d \cos(\omega_d t + \phi_{i}).
\end{align}
We set equal driving phase $\phi_{i=A,B}$ and strength $\eta_i$ in the two axis ($s=\uparrow,\downarrow$) of the micro-traps. 
They may be however different opening further possibilities for Floquet engineering. 

In order to resonantly couple the different vibron types we set $\omega_\uparrow-\omega_\downarrow = n\omega_d$, with $n=1,2,\dots$. 
Assuming that the driving period $2\pi/\omega_d$ is much shorter than the time scale $md_x^3\omega_s/C_0$ of vibron hopping, and employing Floquet formalism~\cite{bermudez2011}, the states $\uparrow$ and $\downarrow$ acquire equal energy in the rotating frame, and are coupled by the effective time-independent hopping matrix $\tilde T$~(see Fig.~\ref{fig:schemes}~(b))~\cite{footnote-SM}: 
%\cite{[{The effective hopping matrix is derived in the Supplementary Material}]SMdrive}:
\begin{align}
\frac{\tilde T_{ss'}(\theta)}{T_{ss'}(\theta)}&=J_{f_{ss'}}(Z_{AB})e^{-if_{ss'}(\phi_A-\gamma_{AB})}
\end{align}
with $f_{ss'} =in\sigma^y_{ss'}$, $\sigma_y$ the Pauli matrix, $J_m(x)$ the $m$-th Bessel function of the first kind, and 
\begin{align}
    Z_{AB}^2&=\eta_A^2+\eta_B^2-2\eta_A\eta_B\cos(\phi_B-\phi_A)\\ \tan\gamma_{AB}&=\frac{\eta_B\sin(\phi_B-\phi_A)}{\eta_A-\eta_B\cos(\phi_B-\phi_A)}.
\end{align}
Note that the elements of the hopping matrix have acquired an effective Peierls phase~\cite{peierls1933}. 
The effective Hamiltonian is then
\begin{align}
H = \hbar \sum_{s,s'} 
\left ( a_{Bs}^\dag \tilde T_{ss'}(\theta) a_{As'} + \mathrm{H. c.} \right ).
\label{eq:drivenH}
\end{align}
The driving can be employed not only to bridge the energy difference between vibron states, but also to dynamically decouple two sites, using the nodes of $J_m(x)$. 
We make use of this flexibility below in order to induce and monitor non-Abelian dynamics in the system.

%%%%%%%%%%%%%%%%%%%%%%%%%%%%%%%%%%%%%%%%%%%%%%%%%%%%%%%%%%%%
%%%PLAQUETTE

\paragraph{Non-Abelian plaquette.--} 
In the following, we employ the technique described above to engineer non-Abelian dynamics in a minimal instance, formed by four micro-traps arranged in a square plaquette on the $xy$-plane, see Fig.~\ref{fig:schemes}~(c). 
However, the ideas can be extrapolated to larger two-dimensional lattices.
We label the ions A, B, C and D in a counter-clockwise order as in Fig.~\ref{fig:schemes}~(c).

The tilting angle of all micro-traps is set to $\theta=\pi/4$, which results in bare vibron hopping matrices along the two directions given by $T_x = T(\pi/4)$ and $T_y=T(-\pi/4)$, where $T(\theta)$ is given by Eq.~\eqref{eq:T}. 
Ions diagonally opposite to each other are coupled with a weaker amplitude, $T_{x+y} =T(0)/2\sqrt{2}$ and $T_{x-y}=T(\pi/2)/2\sqrt{2}$. 
We consider for simplicity the same driving strength $\eta$ in all micro-traps, and the same phase difference $\delta\phi=\pi$ between sites placed at opposite sites of a plaquette diagonal.
Note that with the choice of $\theta=\pi/4$, the hopping matrices $T_{x+y}$ and $T_{x-y}$ are diagonal.

Hence, ions at opposite sites of a diagonal are fully decoupled if $\eta$ is chosen such that $J_0(Z_{AC})=0$ and $J_0(Z_{DB})=0$~(note that restricting the hopping to nearest neighbors simplifies the dynamics, but it is not fully necessary for establishing non-Abelian dynamics).
Due to our choice of $\delta\phi$ this requirement simplifies to $\eta=0.5j_{01}$ with $j_{01}$ the first root of $J_0(x)$~\cite{footnote-SM}.
%\cite{[{For a discussion of the driving parameters in a square see the Supplementary Material}]SMparams}.
Setting $\phi_A=0$, the only free phase parameter left is $\phi=\phi_D$.
We can then write the effective hopping matrices of the driven system as:
\begin{widetext}
\begin{align}
\tilde T_x&=-\frac{C_0}{2md^3}
\begin{pmatrix}
\frac{J_0\left(j_{01}|\cos(\phi/2)|\right)}{\omega_\uparrow} & -3\frac{J_n\left(j_{01}|\cos(\phi/2)|\right)}
{\sqrt{\omega_\uparrow\omega_\downarrow}} e^{-in\phi/2} \\ -3\frac{J_n\left(j_{01}|\cos(\phi/2)|\right)}{\sqrt{\omega_\uparrow\omega_\downarrow}} e^{in\phi/2}& \frac{J_0\left(j_{01}|\cos(\phi/2)|\right)}{\omega_\downarrow}
\end{pmatrix} \\
\tilde T_y&=-\frac{C_0}{2md^3}
\begin{pmatrix}
\frac{J_0\left(j_{01}|\sin(\phi/2)|\right)}{\omega_\uparrow} & 3i^n\frac{J_n\left(j_{01}|\sin(\phi/2)|\right)}
{\sqrt{\omega_\uparrow\omega_\downarrow}} e^{-in\phi/2} \\ 3i^n\frac{J_n\left(j_{01}|\sin(\phi/2)|\right)}{\sqrt{\omega_\uparrow\omega_\downarrow}} e^{in\phi/2}& \frac{J_0\left(j_{01}|\sin(\phi/2)|\right)}{\omega_\downarrow}
\end{pmatrix}.
\end{align}
\end{widetext}
The dynamics acquires a non-Abelian character when $[\tilde T_x, \tilde T_y]\neq 0$. 
%%%%%%%%%%%%%%%%%%%%%%%%%%%%%%%%%%%

% MONITORING THE NON-ABELIAN CHARACTER: MOTION AFFECTS SPIN

\paragraph{Path-dependent spin dynamics.--} 
In order to probe the non-Abelian vibron dynamics, we initialize a single vibron in a fixed spin state in site $A$, and monitor its spin state when it reaches the diagonally opposite site $C$~(see Fig.~\ref{fig:schemes}~(c)). 
Path-dependent spin dynamics is revealed if that spin state depends on whether the vibron travelled in a clockwise or counter-clockwise fashion around the plaquette. 
This may be probed by exploiting the flexibility of the proposed system to steer the vibron motion. 
Setting the trap frequencies of trap $B$~($D$) off-resonant we may suppress the hopping to that site, forcing the vibron into the clockwise~(counter-clockwise) path around the plaquette. 
This may be achieved by increasing the frequencies by $(\omega_\uparrow-\omega_\downarrow)/2$, which places them within the energy gap of the other sites, and out of reach of any hopping process.

\begin{figure}[t!]
\includegraphics[width=0.45\textwidth]{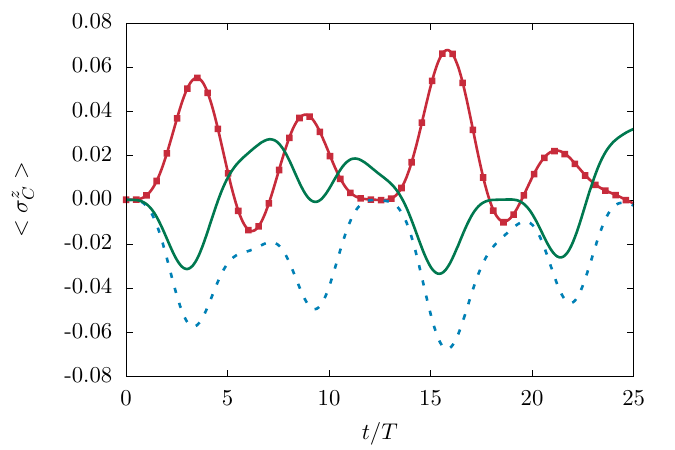}
\caption{
Spin dynamics of ion $C$~(see Fig.~\ref{fig:schemes}~(c)) after initializing a vibron at site $A$ with equal projection to both spin states (see text).
For the red and the blue graph the driving parameters are set to $n=2,~\delta\pi=\pi$ and $\phi=\pi/3$.
The vibron is forced onto a clockwise (red solid line) and a counter-clockwise (blue dashed line), the red squares are results obtained from the time-dependent model without applying the final rotating-wave approximation.
We show the same dynamics for the Abelian case $\phi=\pi/2$ (green solid line).
}
\label{fig:dyn}
\end{figure}

\begin{figure}
    \centering
    \includegraphics[width=0.45\textwidth]{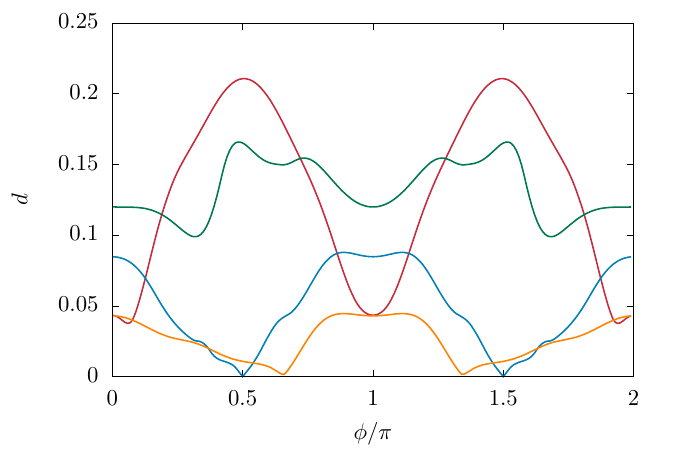}
    \caption{$D_\uparrow$~(see text) as a function of the phase difference $\phi$ of the driving of micro-traps $A$ and $D$~(see Fig.~\ref{fig:schemes}~(c)), for $n=1$ (red), $n=2$ (blue), $n=3$ (green) and $n=4$ (orange). The initial state and the tilting angle $\theta$ are the same as in Fig. \ref{fig:dyn}. The integration time is $T=10 m\omega_\uparrow d^3/C_0$.}
    \label{fig:difintegrate}
\end{figure}

Figure~\ref{fig:dyn} shows the results assuming a single vibron initially prepared at site $A$ in a symmetric superposition of both spin states, $\frac{1}{\sqrt{2}}(a_{A,\uparrow}+a_{A,\downarrow})^\dagger\ket{0}$.
We consider $\omega_d/2\pi=\SI{500}{\kilo\hertz}$, $d=\SI{40}{\micro\metre}$ and $\omega_\uparrow/2\pi=\SI{4}{\mega\hertz}$, within the range reported in recent experiments aiming at two-dimensional quantum simulations with ions~\cite{kiefer2019}.
We have compared our results using the effective time-independent Hamiltonian~\eqref{eq:drivenH}, with those obtained using the time-dependent Hamiltonian finding a good agreement for the considered modulation frequency. 
Driving with $n=2$, $\phi=\pi/2$ and $\delta\phi=\pi$, induces $\tilde T_x=\tilde T_y$, and, as it should, the vibron spin at $C$ remains independent of the travelled path. 
When tuning the driving away from this point, the vibron spin at site $C$ becomes path-dependent revealing the non-commutative dynamics. 
In order to quantify the non-Abelian character, we define:
\begin{align}
D_s=\frac{1}{T}\int_0^T|n_s^{\circlearrowleft}(t)-n_s^\circlearrowright(t)|dt
\end{align}
where $n_s^\circlearrowright(t)$~($n_s^{\circlearrowleft}(t)$) is the vibron occupation of ion $C$ in the state $s=\uparrow,\downarrow$ with the clockwise~(counter-clockwise) path open.
Figure~\ref{fig:difintegrate} depicts $D_\uparrow$ as a function of the driving phase difference $\phi$ and $n$, fixing $\delta\phi=\pi$. 
We observe path-dependence for basically any choice of the driving, showing that non-commutative dynamics is a robust general feature of the set up that requires no precise fine-tuning.
%%%%%%%%%%%%%%%%%%%%%%%%%%%%%%%%%%%%%%%%%%%%%

% Dynamics

% LUIS: For the blue line choose better dotted instead of dashed. It will look better.

%%%%%%%%%%%%%%%%%%%%%%%%%%%%%%
%TIME INTEGRAL
%%%%%%%%%%%%%%%%%%%%%%%%%%%%%%

%%%%%%%%%%%%%%%%%%%%%%%%%%%%%%%%%%%%%%%%%%%%%

% FIGURE 4

%%%%%%%%%%%%%%%%%%%%%%%%%%%%%%%%%%%%%%%%%%%%%

%%%%%%%%%%%%%%%%%%%%%%%%%%%%%%
%ANGULAR MOMENTUM
%%%%%%%%%%%%%%%%%%%%%%%%%%%%%%#

%%%%%%%%%%%%%%%%%%%%%%%%%%%%%%%%%%%%%%%%%%%%%

% FIGURE 5

\begin{figure} [t]
    \centering
    \includegraphics[width=0.45\textwidth]{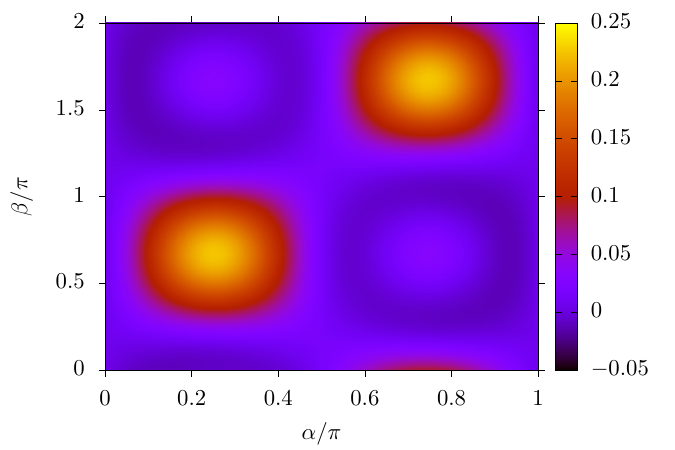}
    \caption{Average angular momentum $\mathcal{L}$ as a function of the initial state given by Eq.~\eqref{eq:initstate}. The driving has been set to $n=2$, $\phi=\pi/3$ and $\delta\phi=\pi$. The integration time $T$ is the same as in Fig.~ \ref{fig:difintegrate}. The value of ${\cal L}$ is normalized by $C_0/m\omega_\uparrow d$.}
    \label{fig:angmom}
\end{figure}

%%%%%%%%%%%%%%%%%%%%%%%%%%%%%%%%%%%%%%%%%%%%%

% MONITORING THE NON-ABELIAN CHARACTER: SPIN AFFECTS MOTION

\paragraph{Spin-orbit coupling.--} 
The spin-orbit coupling characteristic of the non-commutative nature of the vibron dynamics may be alternatively probed by monitoring how the preparation of a specific initial spin state can affect the vibron motion. 
We consider different initial spin states at site $A$:
\begin{align}
    \ket\psi = \left(e^{i\beta}\sin\alpha\, a_{A,\uparrow}^\dagger+\cos\alpha \, a_{A,\downarrow}^\dagger\right)\ket 0, 
    \label{eq:initstate}
\end{align}
and analyze the time-averaged vibron angular momentum 
\begin{align}
\mathcal{L} = \frac{1}{T}\int_0^T dt \vec r(t) \times \vec v(t),
\end{align}
where $\vec r = \sum_j n_j \vec R_j$ is the vibron position, and $\vec v = \sum_j \frac{dn_j}{dt} \vec R_j$ is the vibron velocity, with $n_j=\braket{\sum_s a_{j,s}^\dagger a_{js}}$ the local vibron number. 
Non-Abelian vibron dynamics results in a spin dependent angular momentum, as shown in Fig.~\ref{fig:angmom}, where we depict $\mathcal{L}$ as a function of $\alpha$ and $\beta$ for 
the same case as in Fig.~\ref{fig:dyn}~(b), i.e. $n=2$, $\delta\phi=\pi$, and $\phi=\pi/3$~(the orientation of the rotation is inverted when $\phi\rightarrow -\phi$). 
Note that this way of probing the dynamics merely involves free evolution in the plaquette, without the need of engineering any given path, as above.

The results for the time average of the angular momentum as well as the spin occupation difference presented in the section above depend on the integration time and the driving parameters. 
In particular choosing larger $T$ reduces the observable signal for $\mathcal{L}$ since in any case the angular momentum of the vibron is oscillating around $0$. 
Note in this sense, that unlike a charged particle subjected to a magnetic field, vibrons do not perform a directed cyclotron motion.

%%%%%%%%%%%%%%%%%%%%%%%%%%%%%%%%%%%%%%%%%%%%%%
%% 2D Lattice

\paragraph{Non-Abelian dynamics in a square lattice.--}
\begin{figure}
    \centering
    \includegraphics[width=0.45\textwidth]{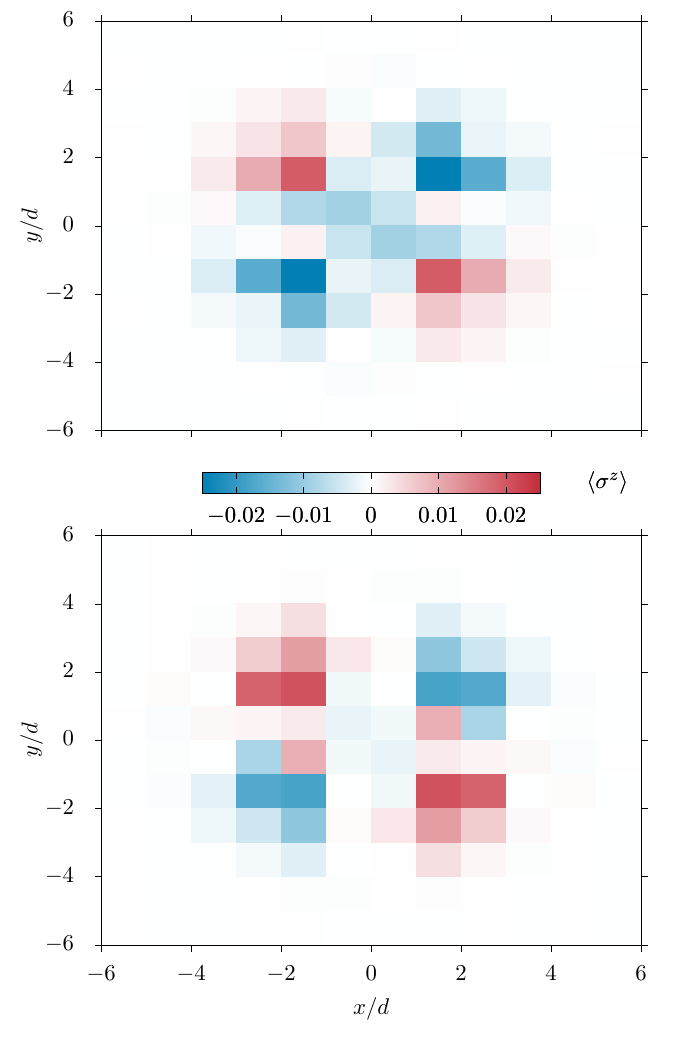}
    \caption{Expectation values $\braket{\sigma_i^z}$ in a square lattice created from copies of the square plaquette shown in Fig.~\ref{fig:schemes}(c). 
    The initial vibron excitation with spin state given by $\alpha=\pi/4,~\beta=3\pi/5$ has an equal projection to the central four sites.
    The evolution time was $2m\omega_\uparrow d^3/C_0$.
    The driving phase has been chosen to $\phi=\pi/3$ (top) and $\phi=\pi/2$ (bottom).}
    \label{fig:LatticeDynamics}
\end{figure}
Finally, we consider a square lattice of microtraps consisting of copies of the square plaquette discussed in the section above.
In this system, vibrons may hop between sites beyond nearest neighbors, with a hopping rate decaying as $1/|\vec r_i-\vec r_j|^3$. 
This decay is partially countered by resonances between the driving phases, e.g. the microtrap frequencies of sites with distance $2d$ are oscillating in phase and hence the Floquet drive enhances the respective hopping rates.
As a consequence we cannot neglect the long-range hopping matrices for the calculation of the vibron dynamics.
In Fig. \ref{fig:LatticeDynamics} we investigate the evolution of a vibron excitation initially delocalized equally over the sites of one square plaquette with a fixed spin state.
The non-Abelian hopping matrices lead to a different diffusion of the two spin components and hence the buildup of spin state separation.
This mechanism is expressed in a breaking of the $\mathbb{Z}_4$ symmetry of the initial state, i.e. the spin state distribution is not invariant under the rotation of the system by multiples of $\pi/2$.

Interestingly, long-range hops, which constitute a relevant feature of the dynamics of driven vibrons in ion lattices, result in non-Abelian dynamics for all possible parameters. 
In this sense, as seen in Fig.~\ref{fig:LatticeDynamics} the choice of $n=2,~\phi=\pi/2$, leading to $\tilde T_x=\tilde T_y$~(and hence to Abelian dynamics in the single-plaquette case), results in non-Abelian dynamics as well, caused by hops over a distance of $\sqrt{5}d$. This is yet another sign of the robust non-Abelian character of the vibron dynamics in the considered driven ion arrays.

%%%%%%%%%%%%%%%%%%%%%%%%%%%%%%%%%%%%%%%%%%%%%%%%%%%%%%%%%%%%%%%%%%%%%%%%%%%%%%%%%%%%%

% CONCLUSIONS

\paragraph{Conclusions and outlook.--}
The exquisite control possibilities offered by ions in micro-trap arrays open interesting perspectives for the quantum simulation of non-Abelian lattice models. 
Vibrational excitations along the two axes of the micro-traps can be understood as effective spin-$1/2$ particles~(vibrons) hopping amongst the micro-traps. 
Vibron hopping may be accompanied by a spin flip, resulting in a $2\times 2$ hopping matrix with off-diagonal terms. 
Crucially, the hopping matrices in different directions may be engineered as non-commuting, resulting in non-Abelian vibron transport.

We have shown that non-commutative vibron transport may be robustly induced, without the need of a precise fine-tuning of the parameters, by combining two main ingredients. 
First, the micro-trap axes are arranged sustaining an angle with the links joining the micro-trap centers. 
Second, Floquet techniques based on the periodic modulation of the trap frequencies allow for the activation of originally off-diagonal couplings. 
Moreover, we have discussed how these state-of-the-art techniques may be employed to probe in various ways the non-Abelian character by monitoring the induced coupling between the spin and motional degrees of freedom. 

The possibilities for a local individual control of the micro-traps open intriguing possibilities for the engineering of more complex non-Abelian lattice models. In particular, note that we have only discussed single-particle dynamics as the excitations considered here are non-interacting oscillator modes. However, the motional states that encode the vibron spin can be coupled to internal electronic states of the ions by means of a side-band laser, resulting in vibron-vibron interactions~\cite{toyoda2013,ohira2021}. The study of non-commutative dynamics in interacting many-body systems constitute an exciting future direction open by the present work.

\bibliographystyle{apsrev4-1}

\bibliography{Non-AbelianPaper.bib}

\end{document}

\title{Supplementary Material for "Non-Abelian vibron dynamics in trapped-ion arrays"}
\author{L. Timm}
\email[]{lars.timm@itp.uni-hannover.de}
\affiliation{Institut f\"ur Theoretische Physik, Leibniz Universit\"at Hannover, Appelstr. 2, 30167 Hannover, Germany}
\author{H. Weimer}
\affiliation{Institut f\"ur Theoretische Physik, Leibniz Universit\"at Hannover, Appelstr. 2, 30167 Hannover, Germany}
\affiliation{Institut f\"ur Theoretische Physik, Technische Universit\"at Berlin, Hardenbergstr. 36,
10623 Berlin, Germany}
\author{L. Santos}
\affiliation{Institut f\"ur Theoretische Physik, Leibniz Universit\"at Hannover, Appelstr. 2, 30167 Hannover, Germany}

\date{\today}
\begin{abstract}
In this Supplementary Material we provide a detailed derivation of the vibron Hubbard Hamiltonian and the Floquet driving discussed in the main text.
\end{abstract}
\maketitle

%%%%%%%%%%%%%%%%

\paragraph{Vibron Hamiltonian. --}
In the following we sketch the derivation of the effective Hamiltonian (Eq.~(9) in the main text) and explicitly discuss the approximations applied to obtain it.
In our work we deal with the quantum vibrations of the ions around their classical equilibrium positions, the vibrons.
If we consider the case of two ions, $A$ and $B$, with mass $m$ confined in microtraps with trap centers $\vec R_{j=A,B}=\pm\frac{d}{2}\vec e_x$ the mutual Coulomb repulsion shifts the classical equilibria away from the trap centers. 

However, for large $d$ and stiff local confinement the classical equilibrium positions are in good approximation given by $\vec R_j$.
The Hamiltonian for the displacement vectors of the ions $\vec r_i$ is then given by Eq.~(1) of the main text.
We expand the Coulomb term up to second order in $|\vec r_A-\vec r_B|/d$ and neglect the first order term since it vanishes for the equilibrium positions as argued above. The remaining term of the expansion can be written in the form 
\begin{align}
    H_{hop}=\frac{1}{2d^3}\left(3[\vec e_x\cdot(\vec r_A-\vec r_B)]^2-[\vec r_A-\vec r_B]^2\right).
\end{align}
Introducing the creation and annihilation operators $a_{js}^\dagger$ and $a_{js}$ we write the displacement operators as $\sqrt{2}\vec r_i=\sum_s l_s\vec e_s\left(a_{is}^\dagger+a_{is}\right)$.
The expansion yields number non-conserving terms which are proportional to $\frac{C_0l_sl_{s'}}{d^3}a_{is}^\dagger a_{js'}^\dagger + \mathrm{H. c.}$. 
In an interaction picture these terms rotate with frequency $\omega_s+\omega_{s'}$ so that we can neglect them if 
\begin{align}
    m\omega_s^2d^2\gg\frac{C_0}{d}.\label{eq:RWAnoncon}
\end{align} 

After this rotating wave approximation the vibron Hamiltonian takes the form
\begin{align}
    H&=\hbar\sum_{j,s}\left(\omega_s+d\omega_s\right)a_{js}^\dagger a_{j,s} \label{eq:HCoul}\\
    &+\hbar g\sum_j\left(a_{j\uparrow}^\dagger a_{j\downarrow}+a_{j\downarrow}^\dagger a_{j\uparrow}\right) \nonumber\\
    &+\hbar\sum_{s,s'}\left(a_{B,s}^\dagger T_{ss'}a_{As'}+\mathrm{H. c.}\right),\nonumber
\end{align}
with $d\omega_s=\frac{C_0l_s^2}{2\hbar d^3}\left[3(\vec e_s \cdot \vec e_x)^2-1\right]$, $g=\frac{3C_0l_\downarrow l_\uparrow}{4\hbar d^3}\sin2\theta$ and with $T_{ss'}$ the vibron hopping matrix given in Eq.~(4) of the main text.
We can neglect the energy shifts of the vibrons $d\omega_s$ since $\omega_s\gg d\omega_s$ due to condition \eqref{eq:RWAnoncon}.
The second term of Eq.~\eqref{eq:HCoul} couples the two local vibron directions with coupling strength $g$, leading to a rotation of the local vibron eigenstates.
Employing again rotating wave arguments we can drop these terms for 
\begin{align}
    \omega_\uparrow-\omega_\downarrow \gg \frac{3C_0}{4m\sqrt{\omega_\uparrow\omega_\downarrow}d^3}\sin2\theta,
\end{align}
so that we arrive at the vibron Hubbard Hamiltonian of Eq.~(3) of the main text.
Note that the last condition is qualitatively different from \eqref{eq:RWAnoncon} as it does not deals with the absolute values of the trap frequencies but with their difference, which needs to be sufficient such that the local vibron eigenstates are imposed by the microtraps.

%%%%%%%%%%%%%%%%

\paragraph{Floquet driving. --}
In the following we derive the time-independent model that applies if the trap frequencies are modulated in time with a sufficiently large frequency $\omega_d$.
After inserting the periodically driven frequencies $\omega_{is}=\omega_s+\eta_i\omega_d\cos(\omega_d t +\phi_i)$ into Eq.~\eqref{eq:HCoul}, we transform the Hamiltonian into the interaction picture with respect to its first term $\hbar\sum_{j,s}\omega_sa_{js}^\dagger a_{js}$.
The transformation yields a time-dependent hopping matrix with elements that can be written via a Jacobi-Anger expansion in the form 
\begin{align}
     \frac{T_{ss'}(\theta,t)}{T_{ss'}(\theta)}&=\sum_{m,m'}J_m(\eta_B)J_{m'}(\eta_A)e^{i\left(\omega_s-\omega_{s'}+(m-m')\omega_d\right)t}\label{eq:timeT} \\
    &\times  e^{i\left(m\phi_B-m'\phi_A\right)}.\nonumber
\end{align}
When using the shorthand notation $\omega_s-\omega_{s'}=f_{ss'}\omega_d$, where $f_{ss'}=in\sigma_{ss'}^y$ is defined as in the main text, we find that terms with $m'=m+f_{ss'}$ are resonant as their time-dependence cancels.
All other terms in the sum oscillate with a frequency that is given by a multiple of $\omega_d$.
These off-resonant terms can be neglected if 
\begin{align}
    \frac{2\pi}{\omega_d}\ll \frac{md^3\omega_s}{2\pi C_0}.
\end{align}
and the timescale of the rotation separates from the timescale of the vibron hopping given by the amplitude of $T_{ss'}(\theta)$.

The sum of the resonant terms is then given by
\begin{align}
\frac{\tilde T_{ss'}(\theta)}{T_{ss'}(\theta)}&=\mathcal{F}(\eta_{B},\eta_{A},f_{ss'},2\delta\phi_{BA}) e^{-if_{ss'}\bar\phi_{BA}}, \label{eq:hopT}
\end{align}
with $2\delta\phi_{ij}=\phi_i-\phi_j$ and $2\bar\phi_{ij}=\phi_i+\phi_j$ and with
\begin{align}
\mathcal{F}(a,b,f,\phi)&=\sum_{m=-\infty}^{\infty} J_m(a)J_{m+f}(b)e^{i(m+f/2)\phi}\\
&=J_f(Z)e^{i(\gamma+\phi/2)f}, \label{eq:F}
\end{align}
where $J_m(x)$ $m$-th Bessel function of the first kind and we used the Neumann-Graf formular in the second line \cite{watson1995}.
This introduces the parameters
\begin{align}
    Z^2=a^2+b^2-2ab\cos\phi,\quad \tan\gamma=\frac{a\sin\phi}{b-a\cos\phi}.
\end{align}
For $a=b$ these parameters simplify to $Z=2|a\sin\frac{\phi}{2}|$ and $\tan\gamma=\cot\frac{\phi}{2}$, which we can solve to $2\gamma=\pi-\phi$ so that the hopping matrix terms are given by
\begin{align}
   \frac{\tilde T_{ss'}(\theta)}{T_{ss'}(\theta)}&=(i)^{f_{ss'}}J_{f_{ss'}}(2|\eta\sin\delta\phi_{BA}|)e^{-if_{ss'}\bar\phi_{BA}}\nonumber.
\end{align}

%%%%%%%%%%%%%%%%%%%

\paragraph{Dynamical decoupling. --}

%%%%%%%%%%%%%%%%%%

% FIGURE S1

\begin{figure}
     \centering
     \includegraphics[width=0.45\textwidth]{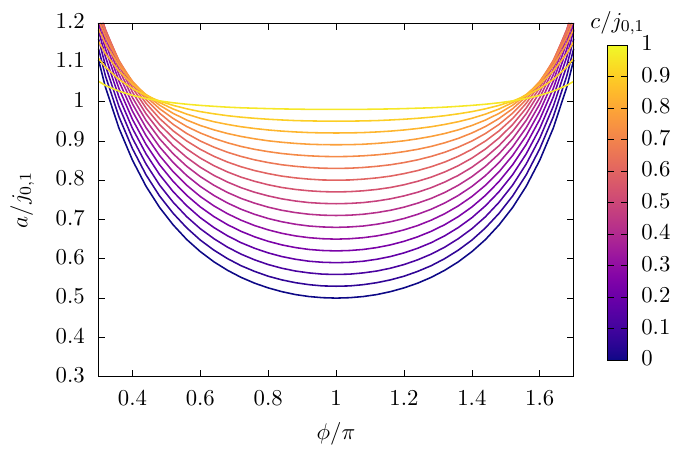}
     \caption{The first roots of the function $|\mathcal{F}(a,a-c,0,\phi)|$ given by Eq.~\eqref{eq:Froots} as a function of $\phi$ for different values of $c$. The driving amplitudes are normalized by the first root of the zeroth Bessel function $j_{0,1}\approx2.4048$.}
     \label{fig:Froots}
\end{figure}
%%%%%%%%%%%%%%%%%%

In the main text we use the modulation in order to decouple the ions which are located in sites diagonally opposite in a square.
This requires to tune the corresponding driving amplitudes and the phase differences such that $|\mathcal{F}(a,b,0,\phi)|$ vanishes.
As can be seen from Eq.~\eqref{eq:F}, this demands $Z^2=j_{0,k}^2$ where $j_{m,k}$ is the $k$-th positive root of $J_m(x)$.
For arbitrary values $a$ and $b$ we introduce $c=a-b$ and find the solutions
\begin{align}
    a=\frac{c}{2}+\frac{1}{2}\sqrt{c^2+\frac{j_{0,k}^2-c^2}{\sin^2\frac{\phi}{2}}}.\label{eq:Froots}
\end{align}
These solutions are shown in Fig.~\ref{fig:Froots} as a function of the angle $\phi$ with different values of $c$, choosing $k=1$.
A special case relevant for this work is $c=0$ which simplifies the solution to 
\begin{align}
    a=\frac{j_{0,k}}{2|\sin\frac{\phi}{2}|}, \label{eq:zeroes}
\end{align} 
which has its minimal value of all solutions for $\phi=\pi$.

%%%%%%%%%%%%%%%%

\paragraph{Hopping matrices in the square plaquette.--}
When the derived model is applied to the square plaquette discussed in the main text, there are are in general $9$ parameters ($4$ driving amplitudes, $4$ driving phases and $n$), however they cannot be chosen arbitrarily in order to engineer non-Abelian dynamics. Let us assume that the driving strengths of sites diagonally opposite are equal such that we have $\eta_A=\eta_C=\eta_1$ and $\eta_B=\eta_D=\eta_2$ for the two diagonals.
In order to ensure that the hopping matrices in $x$ direction do not depend on the $y$ positions of the coupled ions (and vice versa) we need to impose that the respective matrix elements given by Eq.~\eqref{eq:hopT} are equal.
For equal amplitudes we need $Z_{AB}=Z_{DC}$ and $Z_{AD}=Z_{BC}$ which demands 
\begin{align}
    \cos(\phi_B-\phi_A)&=\cos(\phi_C-\phi_D), \\ \cos(\phi_D-\phi_A)&=\cos(\phi_C-\phi_B).\nonumber
\end{align}
An obvious solution to these two equations is $2\delta\phi_{CA}=2\delta\phi_{BD}=\pi$ which also ensures that the respective phases of the matrix elements are equal.

If we in addition decouple the ions diagonally opposite the driving strengths are given by Eq.~\eqref{eq:zeroes}, i.e. $2\eta_1=j_{0k}$ and $2\eta_2=j_{0f}$.
This includes the case of the main text with $k=f=1$ for which all driving strengths are equal.

With these restrictions to the driving parameters the general form of the hopping matrices is given by
\begin{align}
    \frac{\tilde T^x_{ss'}(\theta)}{T_{ss'}(\theta)}&=J_{f_{ss'}}\left(j_{0k}|\cos\delta\phi_{DA}|\right)e^{-if_{ss'}\bar\phi_{DA}},\nonumber\\
    \frac{\tilde T^y_{ss'}(\theta-\frac{\pi}{2})}{T_{ss'}(\theta-\frac{\pi}{2})}&=(i)^{f_{ss'}}J_{f_{ss'}}\left(j_{0k}|\sin\delta\phi_{DA}|\right)e^{-if_{ss'}\bar\phi_{DA}}.\nonumber
\end{align}

\bibliography{SM.bib}